\documentclass[preprint,12pt]{elsarticle}

\makeatletter
\def\ps@pprintTitle{%
  \let\@oddhead\@empty
  \let\@evenhead\@empty
  \def\@oddfoot{\reset@font\hfil\thepage\hfil}
  \let\@evenfoot\@oddfoot
}
\makeatother

\usepackage{graphicx}
\usepackage{amssymb}

\usepackage{eurosym}
\usepackage{xcolor}
\usepackage[colorlinks=true, urlcolor=black,linkcolor=black]{hyperref}
\usepackage{dirtytalk} 

\newcommand{\hide}[1]{}

\newcommand{\curies}[1]{\nolinkurl{#1}}

\newcommand{\issue}[1]{\href{https://github.com/w3c/dxwg/issues/#1}{Issue #1}}
\newcommand{\requirement}[1]{\href{https://www.w3.org/TR/dcat-ucr/\##1}{Req. #1}}
\newcommand{\requirementID}[1]{\href{https://www.w3.org/TR/2019/NOTE-dcat-ucr-20190117/\##1}{[#1]}} 
\newcommand{\usecase}[1]{\href{https://www.w3.org/TR/dcat-ucr/\#ID#1}{ID#1}}
\newcommand{\subsectionR}[1]{\subsection{#1}}
\newcommand{\subsubsectionR}[1]{\subsubsection{#1}}
\newcommand{\claimR}[1]{\textit{#1}}

\usepackage{eurosym}

\begin{document}
\begin{frontmatter}


\title{The W3C Data Catalog Vocabulary, Version 2: Rationale, Design Principles, and Uptake}



\author{Riccardo Albertoni\fnref{label2}} 
\ead{riccardo.albertoni@cnr.it}
\fntext[label2]{Istituto di Matematica Applicata e Tecnologie Informatiche "Enrico Magenes", Consiglio Nazionale delle Ricerche (IMATI-CNR), Genoa, Italy}
\cortext[cor1]{Corresponding Author}

\author{David Browning\fnref{label3}} 
\fntext[label3]{Independent Consultant, formerly refinitiv.com}
\ead{dave@adbrowning.co.uk}
\author{Simon Cox\fnref{label4}} 
\fntext[label4]{Land and Water, CSIRO, Melbourne, Australia}
\ead{simon.cox@csiro.au, simon.j.d.cox@pm.me}
\author{Alejandra N. Gonzalez-Beltran\corref{cor1}\fnref{label5}}
\ead{alejandra.gonzalez-beltran@stfc.ac.uk}
\fntext[label5]{Scientific Computing, Rutherford Appleton Laboratory, Science and Technology Facilities Council, Harwell Campus, OX11 0QX, United Kingdom}
\author{Andrea Perego\fnref{label7}}
\fntext[label7]{European Parliament, Luxembourg}
\ead{andrea.perego@europarl.europa.eu}
 \author{Peter Winstanley\fnref{label6}}
 \ead{peter.winstanley@semanticarts.com}
 \fntext[label6]{Semantic Arts, Fort Collins, CO 80524, USA}

\begin{abstract}
DCAT is an RDF vocabulary designed to facilitate interoperability between data catalogs published on the Web. Since its first release in 2014 as a W3C Recommendation, DCAT has seen a wide adoption across communities and domains, particularly in conjunction with implementing the FAIR data principles (for findable, accessible, interoperable and reusable data). These implementation experiences, besides demonstrating the fitness of DCAT to meet its intended purpose, helped identify existing issues and gaps. Moreover, over the last few years, additional requirements emerged in data catalogs, given the increasing practice of documenting not only datasets but also data services and APIs. This paper illustrates the new version of DCAT, explaining the rationale behind its main revisions and extensions, based on the collected use cases and requirements, and outlines the issues yet to be addressed in future versions of DCAT. 
\end{abstract}


\end{frontmatter}


\section{Introduction}
\label{S:1}


Data has become the most important asset that enables addressing issues ranging from societal challenges, such as pandemics and climate change, to everyday business insights. Thus, data descriptions and data cataloging are fundamental for supporting these data-driven approaches. The last few years have seen an increase in the trend towards Open Data, originally related primarily to \emph{public sector information}, and then with increasing emphasis on facilitating the sharing and re-use of \emph{research data} ---for example, the Research Data Alliance (RDA)\footnote{\url{https://www.rd-alliance.org/} (accessed 10 February 2023)} and funder policies---, as well as an understanding of the importance of metadata ---for example, with the uptake of FAIR data principles \cite{FAIR} for Findable, Accessible, Interoperable and Reusable data. Besides enabling data discovery and re-use, metadata is now also considered crucial to providing all the information necessary to reproduce an experiment---not only in order to verify the research results in scientific studies, but also in cases where data are used in support to policy making and impact assessment in the public sector. In addition, the qualitative and quantitative costs of not providing FAIR data and metadata have been estimated to be really high: an estimated impact of \euro{10.2 bn} for the European economy \cite{european_commission_directorate_general_for_research_and_innovation_cost-benefit_2018}.

The Data Catalog Vocabulary, or DCAT, is a notable contribution to this picture. DCAT is a metadata vocabulary designed to facilitate interoperability between data catalogs published on the Web, irrespective of the domain, community, or platform. Consequently, by using DCAT, data published on the web can be exchanged between systems in an unambiguous manner and with a shared meaning. It was developed following the World Wide Web Consortium (W3C) standardization processes.

Originally developed and hosted at the Digital Enterprise Research Institute (DERI), DCAT was considered by the W3C e-Government Interest Group, and further refined by the Government Linked Data (GLD) Working Group, which published it as a W3C Recommendation in 2014 \cite{vocab-dcat-1}. Since then, it has been adopted and adapted by different parties---a notable example being DCAT-AP \cite{DCAT-AP}, the profile of DCAT being used across Europe as metadata interchange format. 

In this paper, we describe the revision of DCAT, referred to as DCAT 2, which was developed by the W3C Dataset Exchange Working Group (DXWG)\footnote{\url{https://www.w3.org/2017/dxwg/} (accessed 10 February 2023)} in response to a new set of use cases and requirements gathered from implementation experiences with the original version (2014) of the W3C DCAT vocabulary, and new applications that were not considered at that time.  These include the possibility of cataloging other resource types in addition to datasets, such as data services, and of describing relationships between datasets, as well as between datasets and other cataloged resources. Overall, DCAT 2 harmonizes approaches emerging from different communities of usage, extending the core on which profiles can ensure the uniformity of semantics required for a lossless interoperability. 

DCAT 2 was published as a W3C Recommendation in February 2020 \cite{vocab-dcat-2}. This paper complements the formal recommendation, offering insights into the requirements and the process considered in the new version of DCAT.


The paper is organized as follows.
Section \ref{sec:methodology} explains the methodology, detailing the design principles adopted for
the development of DCAT 2.
Section \ref{sec:ucr} gives a brief summary of the requirements that drove the revision. 
Section \ref{sec:model} presents the DCAT model and highlights the features and guidelines introduced in DCAT 2. 
Section \ref{sec:relwork} reviews and discusses contributions in relation to other well-known metadata vocabularies. 
Section \ref{sec:implementation}  discusses the implementation evidence and the uptake of DCAT.
Finally,
Section \ref{sec:conclusions} summarizes the contributions and outlines future activities.

\section{Methodology and Design Principles}
\label{sec:methodology}


The revision of DCAT has been developed by the W3C Data Exchange Working Group (DXWG), which was chartered to maximize interoperability between services such as data catalogs, e-infrastructures, and virtual research environments.\footnote{See the DXWG charter: \url{https://www.w3.org/2017/dxwg/charter} (accessed 10 February 2023)} The revision of DCAT was one of the planned deliverables, together with two other specifications concerning guidelines for the publication of application profiles and profile-based content negotiation.

DXWG worked on DCAT version 2 between May 2017 and January 2020. The group discussions took place in circa 130 teleconferences and four face-to-face meetings, as well as via the DXWG mailing list, issue tracker and GitHub repository. Following the formal W3C process, all these resources are publicly available, including the agenda and minutes of each meeting.\footnote{All these resources are publicly available from the DXWG wiki: \url{https://www.w3.org/2017/dxwg/} (accessed 06 March 2023)}

The efforts of DXWG have focused on fulfilling requirements expressed in a W3C Working Group Note, the {\em Dataset Exchange Use Cases and Requirements} \cite{Pullmann:19:DXWGRequirements}, which documents 51 use cases collected by the working group, and from which the requirements for the revision were identified.  
Beside the use cases and requirements documented in \cite{Pullmann:19:DXWGRequirements}, the working group took into account the feedback received in response to four intermediate versions of the specification, consisting of three public Working Drafts and a Candidate Recommendation, each publicized within relevant communities. 


This paper explicitly refers to requirements and technical design issues to guide interested readers into interlinked working group resources, which deepen the discussion and elucidate the design choices made. 

The paper references to working group resources as follows:
\begin{description}
\item[Issues] All the DCAT issues are documented in the GitHub spa\-ce of the DXWG. The paper cites them in the text by number, e.g., \issue{1009} for \url{https://github.com/w3c/dxwg/issues/1009}.
\item[Requirements] Requirements are documented in \cite{Pullmann:19:DXWGRequirements} and replicated as separated GitHub issues to track discussion and changes triggered by the requirements. The paper refers to them by their handles, also pointing to the related issues when specific discussions need to be referenced. For example, the paper refers to \say{Dereferenceable identifiers [RDID]} by \requirementID{RDID}, and to its related issue available at \url{https://github.com/w3c/dxwg/issues/53} as 
\issue{53}.
\item[Use Cases] Use Cases are documented in  \cite{Pullmann:19:DXWGRequirements}. The paper refers to them by their identifiers. For example, it refers to \say{Modeling service-based data access [ID18]} as \usecase{18}, available at \href{https://www.w3.org/TR/dcat-ucr/\#ID18}{https://www.w3.org/TR/dcat-ucr/\-\#ID18}.
\end{description}

The working group adhered to the following guiding principles designing DCAT 2.

\paragraph{Preservation of the backward compatibility with existing implementations} In designing DCAT 2, the working group strove to minimize the impact on existing implementations. Governmental agencies have already deployed broadly the DCAT standard, and the working group aimed to preserve current implementations by avoiding the need to enforce changes unless strictly necessary. DCAT 2 does not make obsolete any pre-existing terms, and introduces new practices by complementing those already in place. New implementations of, e.g., application profiles are expected to adopt DCAT 2, while the existing implementations will not need to be upgraded unless owners want to use the new features. In particular, current DCAT deployments that do not overlap with the DCAT 2 new features (e.g., data services, time and space properties, qualified relations, packaging) do not need to change anything to remain conformant with DCAT 2. 

\paragraph{Reuse of terms from consolidated metadata vocabularies} D\-CAT 2 incorporates terms from pre-existing vocabularies where stable terms with appropriate semantics could be found. This is consistent with the Data on the Web Best Practice (DWBP) \#15 \say{Use terms from shared vocabularies, preferably standardized ones, to encode data and metadata.}\cite{DWBP}. DCAT reuses terms from Dublin Core \cite{dcterms}, FOAF \cite{foaf}, and PROV-O \cite{PROV-O}, and defines a minimal set of classes and properties of its own.  Informal summary definitions of the externally-defined terms are included in the DCAT vocabulary for convenience, while authoritative definitions are available from the normative references. Changes to definitions in the references, if any, will be expected to take precedence over the summaries given in DCAT. 

\paragraph{Minimization of the ontological commitment} The group strives to minimize the ontological commitment of DCAT 2. From a practical point of view, that implies avoiding over-axioma\-tiza\-tion of DCAT, e.g., by introducing restrictions that might limit the re-usability of DCAT. Moreover, following the DWBP \#16 \say{Choose the right formalization level} \cite{DWBP}, DCAT 2 has removed or relaxed domain and range restrictions for properties (such as those concerning the specification of data themes, keywords, and landing pages). As a rule of thumb, DCAT  delegates to application profiles the burden of setting restrictions or providing guidelines for specific applications and communities.

\paragraph{Balancing normative specification and Open-World Assumption} The specification of DCAT 2 is influenced by common assumptions made in contexts of the Semantic Web and linked data. In particular, DCAT is a metadata schema based upon the \say{Open-World Assumption} (OWA), and it is defined by using the Resource Description Framework (RDF) data model \cite{RDF11-CONCEPTS}. The OWA implies that the metadata schema is not closed, and it can be extended using types and relationships borrowed from other schemas. RDF promotes an inherently machine-actionable approach, where each term in a metadata schema has its own identifier, which can be used to retrieve the term's semantics, and terms from distinct vocabularies can be jointly used. These assumptions have proven to scale on uncoordinated open environments such as the Web, but the flexibility offered by the OWA must be taken into account when dealing with the notion of conformance.  DCAT-compliant catalogs may include additional non-DCAT metadata fields and additional RDF data in the catalog's RDF description. The contents of all metadata fields that are held in the catalog (and that contain data about the catalog itself), as well as the corresponding cataloged resources and distributions, are included in this RDF description, and are expressed using the appropriate classes and properties from DCAT. All classes and properties defined in DCAT are used consistently with the semantics declared in the DCAT Recommendation. Constraints on instances can be provided using shape languages such as ShEx and SHACL \cite{Labra2017,shex,shacl}.

\section{Requirements for DCAT 2}
\label{sec:ucr}

Table \ref{tab:AddressedRequirements} summarizes the requirements addressed by DCAT 2. The following sections present the modeling solution introduced in DCAT 2, which refer to the requirements in the table.

\begin{table}[!ht]
\scriptsize
\begin{tabular}{p{0.36\linewidth}|p{0.56\linewidth}}
  \textbf{Requirement} &
  \textbf{Description} \\ \hline
  Dataset access  \requirementID{RDSA} &
  Provide a way to specify access restrictions for both a dataset and a distribution. \\
  Distribution schema  \requirementID{RDIS} &
   Define a way to include identification of the schema the described data conforms to. \\
  Spatial coverage  \requirementID{RSC} &
  Provide means to specify spatial coverage with geometries. \\
  Temporal coverage  \requirementID{RTC} &
  Allow for specification of the start and/or end date of temporal coverage. \\
  Funding source  \requirementID{RFS} &
  Provide means to describe the funding (amount and source) of a Dataset (or entire Catalog). \\
  Related datasets  \requirementID{RRDS} &
  Ability to represent the different relationships between datasets. \\
  Project relation  \requirementID{RPR} &
  Provide a means to indicate the relation of Datasets to a project. \\
  Dataset publications  \requirementID{RDSP} &
   Provide a way to link publications about a dataset to the dataset. \\
  Dataset type  \requirementID{RDST} &
  Provide a mechanism to indicate the type of data being described and recommend vocabularies to use given the dataset type indicated. \\
  Qualified forms  \requirementID{RQF} &
  Define qualified forms to specify additional attributes of appropriate binary relations (e.g. temporal context). \\
  Loosely-structured catalog [\issue {253}]&
  Provide a best practice for a loosely-structured catalog.\\
  Distribution definition  \requirementID{RDIDF} &
  Revise definition of Distribution. Provide better guidance for data publishers. \\
  Distribution package  \requirementID{RDIP} &
  Define way to specify content of packaged files in a Distribution.\\
  Distribution service  \requirementID{RDISV} &
  Provide a mean to describe that a distribution is provided by a service. \\
  Dereferenceable id  \requirementID{RDID} &
  Encode identifiers as dereferenceable HTTP URIs. \\
  Primary \& alternative id  \requirementID{RIDALT} &
  Provide means to distinguish the primary and alternative (legacy) identifiers. \\
  Identifier type  \requirementID{RIDT} &
  Indicate type of identifier (e.g. prism:doi, bibo:doi, ISBN). \\
  Quality-related info  \requirementID{RDQIF} &
  Define a way to associate quality-related information with Datasets. \\
  Data quality model  \requirementID{RDQM} &
  Identify common modeling patterns for different aspects of data quality based on frequently referenced data quality attributes found in existing standards and practices. \\
  Dataset citation  \requirementID{RDSC} &
  Provide a way to specify information required for data citation (e.g., dataset authors, title, publication year, publisher, persistent identifier). \\
  Entailment of Schema.org  \requirementID{RES} &
  Define schema.org equivalents for DCAT properties to support entailment of Schema.org compliant profiles of DCAT records.
\end{tabular}
\caption{Requirements addressed in DCAT 2 identified by their IDs. The loosely-structured catalog requirement (Issue 253) emerged from the community in form of GitHub issue.}
\label{tab:AddressedRequirements}
\end{table}

\section{DCAT Metadata Schema}
\label{sec:model}

\begin{figure*}[ht]
\includegraphics[width=0.83\linewidth]{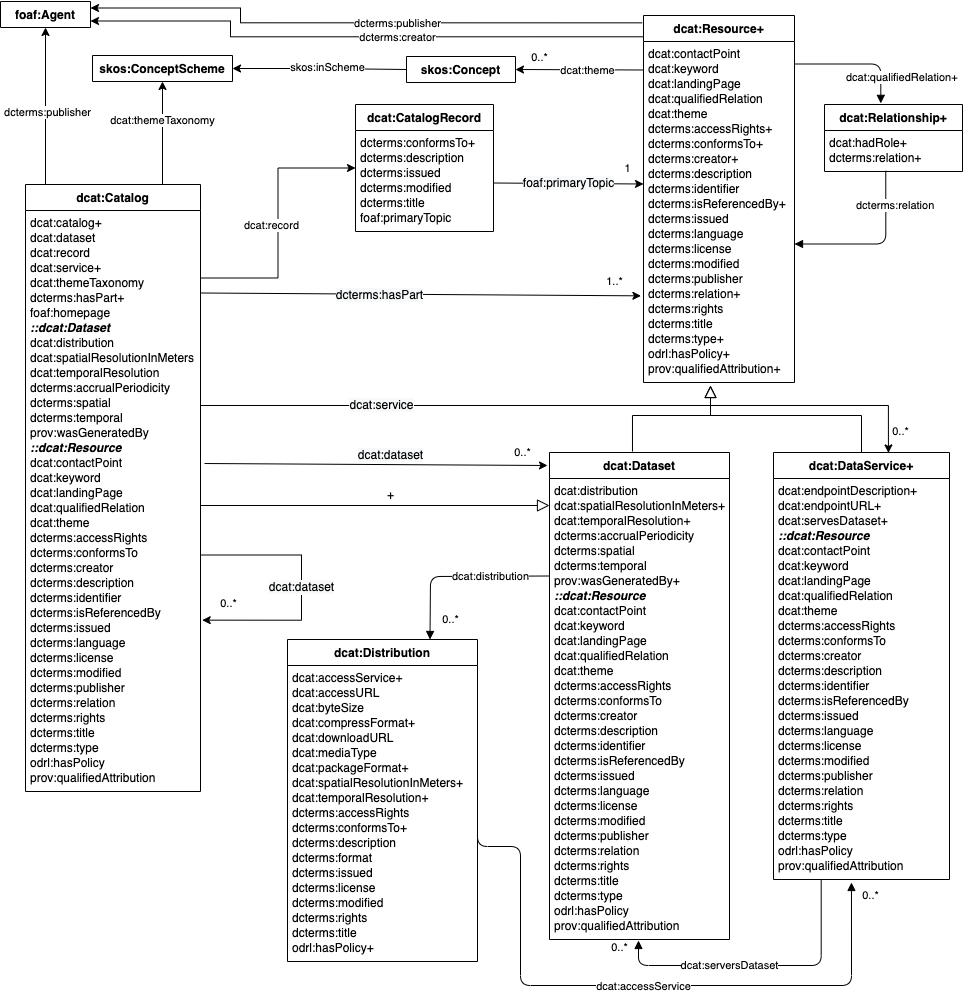}
\caption{Overview of DCAT schema, showing the classes of resources that can be members of a Catalog, and the relationships between them. Classes and terms newly introduced by DCAT 2 are highlighted in the figure by the plus sign.}
\label{fig:DCATDiagram}
\end{figure*}

The backbone of DCAT 2 \cite{vocab-dcat-2} consists of three main classes: \curies{dcat:Catalog}, \curies{dcat:Resource}, \curies{dcat:Distribution}. Figure \ref{fig:DCATDiagram} provides an overview of DCAT 2 model, showing the classes of resources that can be members of a Catalog, and the relationships between them. The diagram uses UML-style class notation, but it should be interpreted following the usual RDF Open-World Assumption around the presence/absence of properties, relationships, and cardinalities.  To assist in understanding the full scope of each class, the inherited properties are copied down from each super-class. Cardinalities are shown in a few places to reinforce expectations, but these are not axiomatized or enforced in any way by the normative recommendation.  

\curies{dcat:Catalog} represents a catalog, which can be seen as a kind of dataset in which each individual item is a metadata record describing a DCAT resource. 
\curies{dcat:Resource} represents any resource that may be described by a metadata record in a catalog. 
It is the parent class of \curies{dcat:Dataset} and \curies{dcat:DataService}--- the most typical  resources types documented in a DCAT catalog. DCAT profiles or applications can define other kinds of resources to be cataloged as sub-classes of \curies{dcat:Dataset}, \curies{dcat:DataService} or \curies{dcat:Resource}. It is worth noting that \curies{dcat:Resource} and its subclasses can be used also for datasets and services which are not included in any catalog. 
\curies{dcat:Distribution} represents a specific representation of a dataset. A dataset might be available in multiple serializations that may differ in various ways, including natural language, media-type or format, schematic organization, temporal and spatial resolution, level of detail or profiles (which might specify any or all of the above).

DCAT 2 borrows from the Dublin Core Metadata Terms (DCTERMS) vocabulary \cite{dcterms} a set of properties that are transversely applicable to different items, including datasets, data services, catalogs, and distributions. In particular,   
\curies{dcterms:title}  and \curies{dcterms:description} to title and describe items; 
\curies{dcterms:issued} and \curies{dcterms:modified} to indicate the date of formal issuance and the most recent modification date of an item; 
\curies{dcterms:license} and \curies{dcterms:rights} to indicate a legal document under which the item is made available and its copyright statements. 

\subsectionR{DCAT 2 new features in the backbone and traversal properties.}
\claimR{DCAT 2 provides guidelines to express conformance}. It recommends the property \curies{dcterms:conformsTo} on a traversal set of items to express conformance to different types of standards. The use of such a property is a consolidated practice in different profiles and vocabularies (e.g., DCAT-AP \cite{DCAT-AP} and DQV \cite{VOCAB-DQV}). Besides, for formal standards issued by bodies like ISO and W3C, \curies{dcterms:conformsTo} is adopted to indicate models, schemas, ontologies, profiles that a cataloged resource or distribution conforms to (see \issue{55} and \issue{411}).

\claimR{DCAT 2 elaborates the guidelines to handle licenses and rights} (see \issue{114}). Different best practices recommend providing data license and right information (e.g. DWBP \cite{DWBP}). However, multiple use cases fall under the umbrella of license and right information. DCAT 2 provides guidelines distinguishing three main cases: one to associate a resource that represents \say{license}; a second, to associate  a resource denoting only access rights (e.g., whether data can be accessed by anyone or just by authorized parties (\requirement{RDSA}, \issue{59})); a third, to cover all the other cases - i.e., statements not concerning licensing conditions and/or access rights (e.g. copyright statements). 

For the first case, DCAT 2 recommends the property \curies{dcterms:license} to refer to canonical URIs of well-known licenses such as those defined by Creative Commons. For the second, it recommends the property \curies{dcterms:accessRights} to express statements specify access rights by referring to code lists/taxonomies, such as the access rights code list MDR-AR\footnote{\url{https://publications.europa.eu/en/web/eu-vocabularies/at-dataset/-/resource/dataset/access-right} (accessed 10 February 2023)} used in DCAT-AP \cite{DCAT-AP} or the Eprints Access Rights Vocabulary Encoding Scheme\footnote{\url{http://www.ukoln.ac.uk/repositories/digirep/index/Eprints\_AccessRights\_Vocabulary\_Encoding\_Scheme} (accessed 06 March 2023)}. For the third, all the other types of rights statements such as copyright statements, which are not covered by \curies{dcterms:license} and \curies{dcterms:accessRights}, DCAT 2 recommends the property \curies{dcterms:rights}. Finally, in the particular case when rights are expressed via Open Digital Right Language (ODRL) policies, DCAT 2 recommends to use the \curies{odrl:hasPolicy} property as the link from the description of the cataloged resource or distribution to the ODRL policy according to the W3C ODRL model \cite{Iannella:18:OIM} and vocabulary \cite{Myles:18:OVE}, in addition to the corresponding DCTERMS property that matches the same ODRL policy type.

The following subsections provide more detailed descriptions of the specific components of DCAT 2.  

\subsection{Resources}
\label{sec:resources}

The class \curies{dcat:Resource} represents a cataloged resource. In previous versions of DCAT, datasets were the only kind of entities in DCAT catalogs. DCAT 2 newly introduces the \curies{dcat:Resource} class, which is an extension point for defining a catalog of any resource. The original \curies{dcat:Dataset} is a sub-class of \curies{dcat:Resource}. Besides properties transversely applicable, the class \curies{dcat:Resource} includes all the properties that were made available in the previous version of DCAT for datasets and might serve for other kinds of resources in DCAT 2. In particular, \curies{dcat:landingPage} indicates a Web page that can be navigated in a Web browser to gain access to the resources, the catalog, a dataset, its distributions and/or additional information.
\curies{dcat:contactPoint}, \curies{dcterms:creator} and \curies{dcterms:publisher} indicate respectively the contact information for the cataloged resource (expressed in vCard \cite{vcard}), the entity responsible for creating the resource and the entity for making the resource available, both expressed as \curies{foaf:Agent}. \curies{dcterms:language} refers to the natural language used for textual metadata (i.e. titles, descriptions, etc) of a cataloged resource.
\curies{dcat:keyword} classifies the resources using free-text keywords, while \curies{dcat:theme} classifies resources with concepts taken from Knowledge Organization Systems (KOS) and possibly available as Linked Data. 

\curies{dcat:Dataset} is a subclass of \curies{dcat:Resource} which represents a collection of data, published or curated by a single agent, and available for access or download in one or more representations, schematic layouts and formats or serializations. The property \curies{dcat:distribution} relates a dataset to its distributions (\curies{dcat:Distribution}). 

\curies{dcat:DataService} is a subclass of \curies{dcat:Resource} which represents a Web API or service that provides access to data, specifically to download distributions of a dataset. 

Other subclasses of \curies{dcat:Resource} can be defined to support applications that catalog other kinds of resource, for example, \say{specimens}. 

\subsubsectionR{DCAT 2 new features in Resource.}
\claimR{DCAT 2 provides flexible mechanisms to indicate the type of cataloged resources} (\requirement{RDST} and \issue{64}). DCAT can be used to model a variety of resources - including documents, software, images and audio-visual content. To ensure the flexibility potentially required by catalogs serving different communities and application cases, DCAT 2  provides two mechanisms for typing resources. First, a cataloged resource description has an RDF type to denote a sub-class of \curies{dcat:Resource} - initially \curies{dcat:Dataset} and \curies{dcat:DataService}. Second, the property \curies{dcterms:type} may be used to indicate a sub-type. It is strongly recommended that the value of this property is taken from a well-governed and broadly recognized set of resource types (e.g., the DCMI Type vocabulary \cite{dcterms}, the DataCite resource types \cite{DataCite}, the ISO-19115-1 scope codes \cite{ISO-19115-1}, the MARC intellectual resource types). Using \curies{dcterms:type} is particularly appropriated for referring to classifications provided by other standards, and to enable interoperability with existing catalogs (see use cases \usecase{8} and \usecase{20}). 
When describing a resource which is not a \curies{dcat:Dataset} or \curies{dcat:DataService}, it is recommended to create a suitable sub-class of \curies{dcat:Resource}, or use \curies{dcat:Resource} with the \curies{dcterms:type} property to indicate the specific type. 

\claimR{DCAT 2 provides information required for data citation} (see \requirement{RDSC} and  \issue{61}). DCAT 2 provides equivalents to all the mandatory elements in DataCite \cite{DataCite}.  The original DCAT already supported  title, publisher, publication year, resource type, DCAT 2 has specifically considered \curies{dcterms:creator} to indicate creator and it provides guidelines for dealing with different types of identifiers (see section \ref{sec:guidelines}).

\claimR{DCAT 2 provides a way to deal with a wide set of relations}.   
Resources might be related in many different ways and complex relations might characterize the context in which resources have been created, for example, to track its input data, the software used,  the agents and founders involved (e.g., see use cases \usecase{9}, \usecase{12}, \usecase{31}, \usecase{32}).
The property \curies{dcterms:relation} is recommended for use in the context of a cataloged resource to capture general relationships, including related datasets (\requirement{RRDS}) and the case where the package of resources associated with a cataloged item includes a mixture of representations, parts, documentations and other elements which are not strictly `distributions' of a dataset (see \issue{253} expressing the requirement on loosely-structured catalogs). 
 The property \curies{dcterms:relation} is a super-property of a number of more specific properties which express more precise relationships, such as  \curies{dcat:distribution}, \curies{dcterms:hasPart}, (and its sub-properties \curies{dcat:catalog}, \curies{dcat:dataset}, \curies{dcat:service}), \curies{dcterms:isPartOf}, \curies{dcterms:conformsTo}, \curies{dcterms:isFormatOf}, \curies{dcterms:hasFormat}, \curies{dcterms:isVersionOf}, \curies{dcterms:hasVersion}, \curies{dcterms:replaces}, \curies{dcterms:isReplacedBy}, \curies{dcterms:references}, \curies{dcterms:isReferencedBy}, \curies{dcterms:requires}, \curies{dcterms:isRequiredBy}.  The  \curies{dcterms:relation} is not inconsistent with a subsequent reclassification with more specific semantics, though the more specialized sub-properties should be used to link a dataset to component and supplementary resources if possible.
For example, DCAT 2 uses the property \curies{dcterms:isReferencedBy} to associate the resource described in the catalog with an external resource that references, cites, or points to the cataloged resource. By applying this property, DCAT 2 tracks publications that reuse or describe a specific dataset (see \requirement{RDSP} and \issue{63}). DCAT 2 tracks the project that has generated a resource: \curies{prov:wasGeneratedBy} links datasets to the projects that have generated them (\requirement{RPR} and \issue{77}). 

\claimR{DCAT 2 supports complex non-binary relations}. It uses qualified relations to deal with relations not covered by the above or other known properties (e.g., PROV-O properties such as \curies{prov:wasDerivedFrom}, \curies{prov:hadPrimarySource}) and to overcome the limitation related to binary relations (see the requirement \say{qualified forms} [\requirement{RQF}] discussed in \issue{79}). Even when the relations are represented in known properties, there may be the need of providing additional information concerning, e.g., the temporal context of a relationship, which requires the use of a more sophisticated representation, for example, to specify the temporal dimension of a role---i.e., the time frame during which an individual/organization played a given role - and, maybe, also other information – e.g., the organization where the individual held a given position while playing that role (see use cases \usecase{19} and \usecase{13}, and \issue{66}). DCAT 2 models relationships between resources and agents with property \curies{prov:qualifiedAttribution} (for example, the funding source \requirement{RFS}) and relationships between resources with \curies{dcat:qualifiedRelation}. Property \curies{prov:qualifiedAttribution} links the resource to instances of the class \curies{prov:Attribution}, which ascribes the resource to an agent indicated by the property \curies{prov:agent}. Property \curies{dcat:qualifiedRelation} links the resource to a relation \curies{dcat:Relationship} involving another  resource  pointed by the property \curies{dcterms:relation}.
The property \curies{dcat:hadRole} is used in  \curies{prov:qualifiedAttribution} to denote the relation the resources have and in \curies{dcat:qualifiedRelation} to indicate the roles an agent plays. 

\claimR{DCAT 2 supports a rich set of temporal and spatial properties to characterize datasets}. The previous version of DCAT offered  \curies{dcterms:issued}, \curies{dcterms:modified} and \curies{dcterms:accrualPeriodicity} to indicates when a dataset is issued, modified and its update schedule.   DCAT 2 adopts new properties specifically dealing with the temporal coverage (\requirement{RTC}).  It introduces the property \curies{dcat:temporalResolution} to specify the minimum temporal separation of items in a dataset encoded as \curies{xsd:duration} and adopts \curies{dcterms:temporal} to indicate the temporal extent of a dataset. The extent is expressed as instances of the class \curies{dcterms:PeriodOfTime}, indicating the start and end of the interval by using properties \curies{dcat:startDate} or \curies{time:hasBeginning}, and \curies{dcat:endDate} or \curies{time:hasEnd}, respectively. The interval can also be open - i.e., it can have just a start or just an end (see \issue{85} for further discussions).
Similarly, DCAT 2 introduces two new properties to express spatial coverage (\requirement{RSC}, see  \issue{83} for the detailed discussion). \curies{dcat:spatialResolutionInMeters} specifies the minimum spatial separation of items in a dataset, expressing it as a decimal values in meters. \curies{dcterms:spatial} expresses the spatial extent of a dataset. Its values are a spatial region or named placed \curies{dcterms:Location}, in which, the property \curies{locn:geometry} specifies an extensive geometry (i.e., a set of coordinates denoting the vertices of the relevant geographic area), \curies{dcat:bbox} specifies
a geographic bounding box delimiting a spatial area, \curies{dcat:centroid} indicates a geographic center of a spatial area, or another characteristic point.

\claimR{DCAT 2 adds mechanisms for including data services}. Data is often served via web services. A service may provide access to more than one dataset, and it is necessary to know how to query the service API to get the data (see use cases \usecase{18} and \usecase{6}). DCAT 2 specializes \curies{dcat:Resource} with a new class \curies{dcat:DataService} to model data services (see \issue{180}). A data service is a collection of operations that provides access to one or more datasets or to data processing. The \curies{dcat:servesDataset} property links a service to data that it can distribute. The kind of service can be indicated using the \curies{dcterms:type} property; its value may be taken from a controlled vocabulary such as the INSPIRE spatial data service type code list\footnote{\url{http://inspire.ec.europa.eu/metadata-codelist/SpatialDataServiceType/} (accessed 06 March 2023)}. \curies{dcat:endpointURL} provides the root location or primary endpoint of the service (a Web-resolvable IRI). Property \curies{dcat:endpointDescription} provides a description of the services available via the endpoints, including their operations, parameters, etc. The endpoint description gives specific details of the actual endpoint instances, using \curies{dcterms:conformsTo} to indicate the general standard or specification that the endpoints implement. An endpoint description may be expressed in a machine-readable form, such as an Open API \cite{OpenApi} description, an OGC GetCapabilities response WFS \cite{WFS,ISO-19142}, WMS \cite{WMS,ISO-19128}, a SPARQL Service Description \cite{SPARQL11-SERVICE-DESCRIPTION}, an OpenSearch \cite{OpenSearch} or WSDL \cite{WSDL20} document, a Hydra API description HYDRA \cite{HYDRA}. 

\subsection{Distributions}
\label{sec:distributions}

\curies{dcat:Distribution} is a specific class for representation of a dataset. A dataset might be available in multiple serializations that may differ in various ways, including natural language, media-type or format, schematic organization, temporal and spatial resolution, level of detail or profiles (which might specify any or all of the above). Distributions represent a general availability of a dataset, whose access can include different access methods (e.g.,  direct download, API, or through a Web Page).
For the distributions, \curies{dcat:downloadURL} provides the URL for a downloadable file in a given format. The \say{format} of a distribution should be specified through the property \curies{dcat:mediaType} when a correspondent IANA Media Types \cite{IANAMediaTypes}  exists, or  \curies{dcterms:format} otherwise.
\curies{dcat:byteSize} specifies the size of distribution in bytes. When a direct link to the downloadable file is not available,  \curies{dcat:accessURL} indicates a URL of the resource that gives access to a distribution of the dataset. It should be used for the URL of a service or location that can provide access to this distribution, typically through a Web form, query or API call. 

\subsubsectionR{DCAT 2 new features in Distributions.} 
\claimR{DCAT 2 introduces distribution service} to support use cases where the distribution of a dataset is made by Web services (\usecase{6} and \requirement{RDISV}).  DCAT 2 adds the property \curies{dcat:accessService} which relates distributions to their \curies{dcat:DataService} detailed information about how users can interact with distribution services (\issue{267}). 

\claimR{DCAT 2 revises and clarifies the definition of distribution} (\requirement{RDIDF}). The previous definition of \curies{dcat:Distribution} allowed a number of alternative interpretations. The definition has been rephrased to clarify that distributions are primarily representations of datasets. DCAT 2 clarifies that lossless transformations between representations are not always possible. In some cases, distributions of the same dataset might have different levels of fidelity to the underlying data (see discussion in \issue{52}). Moreover, the question of whether different representations can be understood to be distributions of the same dataset, or distributions of different datasets, is application-specific. Judgment about how to describe them is the responsibility of the provider, taking into account their understanding of the expectations of users, and practices in the relevant community.\\
\claimR{DCAT 2 supports packaged and compressed distributions} (\requirement{RDIP} see \issue{54}). Distributions can include multiple files made available in compressed archives. DCAT 2 introduces the property \curies{dcat:packageFormat} and \curies{dcat:compressFormat} to indicate the package  and compression formats of the distribution. Both formats should be expressed using a media type as defined by IANA \cite{IANAMediaTypes}, if available.
\claimR{DCAT 2 recommends to indicate distribution schema}. It uses the property \curies{dcterms:conformsTo} to indicate the model or schema used for the representation of dataset (\requirement{RDIS} and \issue{55}).

\subsection{Catalog and Catalog Record}
\label{sec:catalog}

A \curies{dcat:Catalog} is a curated collection of metadata about resources such as datasets and data services.  \curies{dcat:Catalog} is characterized by further properties besides those transversely applicable:   \curies{foaf:homepage} indicates the homepage of the catalog which usually is a public Web document available in HTML;  
\curies{dcat:themeTaxonomy} refers to the Knowledge Organization System (KOS) providing concepts to classify the cataloged resources; 
\curies{dcat:record} links a catalog to a \curies{dcat:CatalogRecord} describing the registration of a single cataloged resource that is part of the catalog. Using  \curies{dcat:record} and \curies{dcat:CatalogRecord} is possible to distinguish between the metadata of a cataloged resource (i.e., instances of \curies{dcat:Resources}) and the metadata of the metadata of the cataloged resource (i.e., instances of \curies{dcat:CatalogRecord}). This is required in specific cases, for example, to express the date when a resource has been registered or modified in the catalog (\curies{dcterms:issued} and \curies{dcterms:modified} attributed to instances of \curies{dcat:CatalogRecord}), which may differ from the publication or modification of the concrete resources (aka \curies{dcterms:issued} or \curies{dcterms:modified} attributed to instances of \curies{dcat:Resource}). 

\subsubsectionR{DCAT 2 new features in Catalog and Catalog Record.}
\claimR{DCAT 2 clarifies the scope of DCAT catalogs}. DCAT was originally conceived to model data catalogs. DCAT 2 opens to novel  first-class cataloged resources providing \curies{dcat:Resource} as an extension point for community-specified cataloged resources (see \issue{172} and section \ref{sec:resources}).
It adds \curies{dcat:DataService} for representing data services and subsumes  \curies{dcat:Dataset} and \curies{dcat:DataService} with \curies{dcat:Resource}.
It provides properties to deal with the new kinds of cataloged resources (see \issue{116}):  \curies{dcterms:hasPart}, to specify a cataloged resource irrespective of its type; \curies{dcat:service}, to specify a cataloged data service. 

\claimR{DCAT 2 enables provision for catalogs to be composed of other catalogs}, in particular, \curies{dcat:Catalog} has been made a sub-class of \curies{dcat:Dataset}, and the property \curies{dcat:catalog} is provided to specify sub-catalogs (see \issue{182}). 

\claimR{DCAT 2 extends the type of thematic resources which can be considered to classify datasets}. It relaxes the global range of the property \curies{dcat:themeTaxonomy} allowing the linking to a KOS that is not formalized as a \curies{skos:ConceptScheme} (See \issue{119}). Beside SKOS concept schemes, SKOS collections \cite{DBLP:journals/ws/BakerBIMSS13, Bechhofer:09:SSK} or OWL ontologies \cite{McGuinness:12:OWO} are recommended advising that each member of the KOS can be denoted by an IRI and published as linked data.  

\claimR{DCAT 2 includes specific mechanisms to state the conformance of metadata to standards}. It adopts the property \curies{dcterms:conformsTo} for \curies{dcat:CatalogRecord} to represent  the conformance of a record metadata with a metadata standard (see \issue{502}). 

\subsection{Guidelines}
\label{sec:guidelines}

In addition to the feature discussed above, DCAT 2 elaborates guidelines to meet specific requirements posed by the community.
Guidelines systematize emerging solutions based on W3C vocabularies such as DQV \cite{VOCAB-DQV} and ADMS \cite{VOCAB-ADMS} which are stable enough to be adopted even if they have not reached the status of W3C recommendation. 

\claimR{DCAT 2 provides guidelines to deal with different kinds of identifiers.} As pointed out in the use case \usecase{11},
a number of different (possibly persistent) identifiers are widely used in the scientific community, especially for publications, but now increasingly for authors and data. Different approaches are used for representing them,  best practices are needed to enable their effective use across platforms. But more importantly, they need to be made actionable, irrespective of the platforms they are used in (see \requirement{RDID}).
Encoding identifiers as HTTP URIs seems to be the most effective way of making them actionable. Notably, quite a few identifier schemes can be encoded as dereferenceable HTTP URIs, and some of them are also returning machine-readable metadata (e.g., DOIs, ORCIDs). Moreover, they can still be encoded as literals, especially if there is the need of knowing the identifier “type” (\requirement{RIDT}). In such a case, a common identifier type registry would ensure interoperability. DCAT 2 reuses terms provided by DCTERMS \cite{dcterms} and VOCAB-ADMS \cite{VOCAB-ADMS}. Data providers can apply \curies{dcterms:identifier} to any kind of resources binding their  HTTP dereferenceable proxy IDs with legacy identifiers, non-HTTP dereferenceable identifiers, locally minted or third-party-provided identifiers (\issue{53}). 
Another issue concerns the ability to specify primary and secondary identifiers. This may be a requirement when resources are associated with multiple identifiers (\requirement{RIDALT}). The property \curies{adms:identifier}  can express other locally minted identifiers or external identifiers, like DOI, ELI, arXiv for creative works, and ORCID, VIAF, ISNI for actors such as authors and publishers, as long as the identifiers are globally unique and stable. The property \curies{adms:identifier} ranges in instances of the class \curies{adms:Identifier}, for which \curies{skos:notation} indicate the identifier as a literal with datatype IRI (e.g.,\verb+"PA 1-060-815"^^ex:type+), \curies{adms:schemaAgency} and \curies{dcterms:creator} represent the authority that defines the identifier scheme (e.g., the ex:type in the example). \curies{adms:schemaAgency} is used when the authority has no URI associated (see \issue{67}). The type of identifiers can be provided as RDF datatypes \cite{RDF11-CONCEPTS} or custom OWL datatypes \cite{OWL2-SYNTAX} if not already registered as URI type. 
Examples of common types for identifier scheme (arXiv, etc.) are defined in DataCite schema\footnote{\url{https://schema.datacite.org/meta/kernel-4.1/include/datacite-relatedIdentifierType-v4.xsd} (accessed 6 March 2023)} 
and FAIRsharing Registry\footnote{\url{https://fairsharing.org/search?q=identifier} (accessed 6 March 2023)} (see \issue{68}).

\claimR{DCAT 2 provides guidelines for documenting the quality of resources and distributions}. Consistently with the recommendations from the Data on the Web Best Practices (DWBP) \cite{DWBP}, the use cases \usecase{45} and \usecase{14} stress the need for a uniform representation of data quality so that consumers understand the possibilities and risks of using and reusing the data. 
DCAT 2 reuses the Data Quality Vocabulary (DQV) \cite{AlbertoniDQV}
\cite{VOCAB-DQV} to associate quality-related information to datasets (\requirement{RDQIF}) and offer common modeling patterns for different aspects of Data Quality (see, \requirement{RDQM}, \issue{57}, \issue{58}). The property \curies{dqv:hasQualityAnnotation} relates datasets and distributions with reviews, users' feedback and quality certificates (modeled as \curies{dqv:QualityAnnotation}). The property \curies{dqv:hasQualityMeasurement}  relates resources and distributions to quality measurements (instances of  \curies{dqv:QualityMeasurement}) evaluated by community-defined domain-specific metrics (\curies{dqv:Metric}) which provide quantitative or qualitative information about the dataset or distribution. \curies{dqv:QualityPolicy} models policies or agreements that are chiefly governed by data quality concerns.   As previously discussed, \curies{dcterms:conformTo} can state the compliance with standards, specifications. DCAT 2 includes examples of how DQV can express the degree of conformance to best practices (e.g. the DWBP \cite{DWBP} or the FAIR Principles \cite{FAIR}) and combines DQV with the Evaluation and Report Language (EARL) \cite{EARL} and PROV ontology \cite{PROV-O} to express details about the results of conformance and quality tests.



\section{Related Work}
\label{sec:relwork}

This section reviews metadata models that readers might perceive as overlapping with DCAT in terms of coverage or goals. The discussion points out the distinct metadata models' peculiarities and their mapping into DCAT. Overall, the discussion clarifies that DCAT is not redundant with the existing metadata models. Instead, a joint of the discussed metadata models with DCAT brings advantages in the overall metadata expressivity and cross-sector, cross-platform sharing, and reuse.



\paragraph{CERIF} The  Common European Research Information Format (CERIF) models Research Environment, including research outputs, persons, organizations, projects, funding programs, facilities as first-class citizens and capturing the semantic relationships of entities with each other as well as entity classifications (i.e. roles). The European Commission mandated euroCRIS to maintain, develop and promote CERIF as an EU recommendation to Member States. euroCRIS now has more than 100 institutional members in approximately 40 countries and there are hundreds of implementations of CERIF, including by several commercial ICT suppliers. CERIF is currently being used in numerous systems in production across Europe (e.g., national or institutional research information systems), as well as in European FP7 e-infrastructure projects, such as OpenAIREplus, {EuroRIs-Net+} and ENGAGE   \cite{DBLP:journals/ijmso/JefferyHJA14}.
CERIF and DCAT differ in terms of goals and specificity.
CERIF specifically focuses on research environments, while DCAT focuses on Data Catalogs. Partial mapping of DCAT into CERIF exists \cite{theodoridou_maria_2019_2548732}. For example, DCAT Datasets can be modeled as \curies{ResultProduct}, but CERIF does not natively provide distinctions between catalogs, datasets, distributions, nor other details such as access details. 

\paragraph{DataCite} The DataCite metadata schema \cite{DataCite} is a list of core metadata properties chosen for accurate and consistent identification of a resource for citation and retrieval purposes, along with recommended use instructions. It is managed by the DataCite consortium,  founded in late 2009 with the goal of easing the access to scientific research data on the Internet, increasing acceptance of research data as legitimate, citable contributions to the scientific record, and supporting data archiving that will permit results to be verified and re-purposed for future study.
DataCite infrastructure is responsible for issuing persistent identifiers (in particular, DOIs) for datasets, and for registering dataset metadata. Such metadata is to be provided according to the DataCite metadata schema. 
While DataCite’s Metadata Schema has been expanded with each new version, it is, nevertheless, intended to be generic to the broadest range of research datasets, rather than customized to the needs of any particular discipline. DataCite metadata primarily supports citation and discovery of data; It does not include specific terms for Catalogs and Distributions,  it is not intended to supplant or replace community-specific metadata.
DataCite enables providing other metadata schemas via DOI content negotiation. In particular, it supports JSON-LD \cite{json-ld11} to serve metadata according to Schema.org. A mapping from DataCite to DCAT is defined in CiteDCAT-AP \cite{CiteDCAT-AP}, a metadata profile used in Zenodo\footnote{\url{https://zenodo.org/} (accessed 06 March 2023)}, the most popular European research data repository.


\paragraph{ISO 19115} ISO 19115-1:2014 \cite{ISO-19115-1} defines a metadata schema for describing geographic information and services by means of metadata. It provides information about the identification, the extent, the quality, the spatial and temporal aspects, the content, the spatial reference, the portrayal, distribution, and other properties of digital geographic data and services. Mapping of ISO 19115 to DCAT has been developed, in particular,  GeoDCAT-AP \cite{GeoDCAT-AP2} is an extension to the “DCAT application profile for European data portals” (DCAT-AP) for the representation of geographic metadata. GeoDCAT-AP was designed to enable the cross-sector and cross-platform sharing and re-use of INSPIRE and, more in general, metadata following the ISO 19115/19119 standards and the corresponding XML-based implementation (ISO 19139).

\paragraph{Schema.org} In 2011, the major search engines Bing, Google, and Yahoo (later joined by Yandex) created Schema.org to provide a single schema across a wide range of topics that included people, places, events, products, offers, and so on \cite{DBLP:journals/cacm/GuhaBM16}. Schema.org is a collaborative, community activity with a mission to create, maintain, and promote schemas for structured data on the Internet, on Web pages, in email messages, and beyond. 
Schema.org includes a number of types and properties based on the original DCAT work (see \curies{sdo:Dataset} as a starting point), and the index for Google's Dataset Search service relies on structured description in Web pages about datasets using both Schema.org and DCAT \cite{DBLP:conf/www/BrickleyBN19}. 
This class is modeled starting from  W3C DCAT work, and benefits from collaboration around the DCAT, ADMS and VoID vocabularies\footnote{See \url{http://www.w3.org/wiki/WebSchemas/Datasets} (accessed 06 March 2023) for full details and mappings.}. 
In particular, Schema.org mimics the DCAT backbone, the  (abstract) \curies{sdo:Dataset} and (concrete) \curies{sdo:DataDownload} matches \curies{dcat:Dataset} / \curies{dcat:Distribution}, as for the relationship of Datasets to DataCatalogs. Contrary to DCAT, Schema.org is not a W3C standard, 
the project is not governed by W3C, the W3C advisory group or the W3C Process; rather, it stems from an informal collaboration.  
In terms of workflow, the primary difference between Schema.org and W3C's recommendation track process is an emphasis on incremental publication of releases (several releases per year) approved by a small steering group whose role is to evaluate and approve release candidates prepared by the project webmaster on the basis of wider discussion which takes place in a dedicated W3C community group and related GitHub project.
DCAT 2 \cite{vocab-dcat-2} provides a mapping between DCAT and Schema.org to clarify the relation between DCAT and Schema.org and promote the discoverability by mainstream search engines (see \requirement{RES}). 

\paragraph{VoID} VoID \cite{void} is an RDF vocabulary for expressing metadata about RDF datasets. 
It covers (i) general metadata following the Dublin Core model; (ii) access metadata describing how RDF data can be accessed using various protocols;  (iii) structural metadata describing the structure and schema of datasets for tasks such as querying and data integration; (iv) description of links between datasets for understanding how multiple datasets are related and can be used together.  VoID is quite popular in the context of Linked data and extended by other vocabularies such as DataID \cite{DBLP:conf/mtsr/FreudenbergBRUE16}. However, 
being specifically suited for RDF dataset and linked data practices, it does not cover all the types of data required by the open and research data community (e.g., CSV, JSON). Fruitfully jointly use of DCAT and VoID have been shown (e.g., by DataID \cite{DBLP:conf/mtsr/FreudenbergBRUE16}).

\section{DCAT implementations and uptake}
\label{sec:implementation}

The W3C recommendation process requires the collection of implementation experiences to show that a specification is sufficiently clear, complete, relevant to market needs, and to ensure that independent, interoperable implementations of each feature of the specification are realized. In view of that, the editors of DCAT 2 prepared a DCAT 2 implementation report \cite{DCAT2-implementations}. The report also shows preliminary evidences of DCAT 2 uptake. It focuses on two types of evidence: i) DCAT-based vocabularies; ii) data catalogs, data services, and datasets.

As for DCAT-based vocabularies, different profiles are based on DCAT 2 \cite{vocab-dcat-2} or extend the original version of DCAT \cite{vocab-dcat-1} with properties and classes included in DCAT 2, showing implementation evidences of the reviews included. Due to the large number of DCAT-based vocabularies and data catalogs supporting DCAT, this section includes only a representative subset, providing nonetheless enough implementation evidence of the revisions proposed in DCAT 2.

In particular, DCAT-AP \cite{DCAT-AP} is a profile of DCAT used across Europe since 2014 as a metadata interchange format, primarily for catalogs of government data, and, to some extent, for scientific data. As such, it has a broad geographic coverage, and it is supported in data catalogs (e.g., the European Data Portal\footnote{\url{https://data.europa.eu/} (accessed 06 March 2023)}) and catalog platforms (e.g., CKAN\footnote{\url{https://ckan.org/} (accessed 06 March 2023)}). 

GeoDCAT-AP \cite{GeoDCAT-AP2} and StatDCAT-AP \cite{StatDCAT-AP} are domain-spe\-cific extensions of DCAT-AP for geospatial and statistical data, respectively, and they share the same geographic coverage of DCAT.

CiteDCAT-AP \cite{CiteDCAT-AP} and DCAT-AP-JRC \cite{DCAT-AP-JRC} are extensions of DCAT-AP specifically designed for multidisciplinary research data, and they are implemented in the corporate catalog of the European Commission's Joint Research Centre\footnote{\url{https://data.jrc.ec.europa.eu/} (accessed 06 March 2023)}. Moreover, CiteDCAT-AP is supported in Zenodo\footnote{\url{https://zenodo.org/} (accessed 06 March 2023)}, the research data catalog and repository most widely used in Europe. 

DCAT-AP has also been used as a basis for the development of country-specific extensions (see \cite{DCAT-AP-EXT}). Such extensions have not been included in this review, but they provide additional support to the implementation evidence for the revisions proposed in DCAT 2 already included in DCAT-AP.

DCAT-AP aligns with DCAT 2 since version 2.0, and such alignment will eventually be reflected in the DCAT-AP extensions. For example, Geo\-DCAT\-AP 2.0 \cite{GeoDCAT-AP2} (released in December 2020) is aligned with DCAT 2.

Moreover, in the context of scientific data, projects and initiatives such as EOSC-pillar \cite{galeazzi_fulvio_2020_4288472}, FAIRsFAIR \cite{devaraju_anusuriya_2020_4081213} and ExPaNDS encourage data repository owners to publish their datasets by mapping their metadata with the DCAT standard when following the FAIR principles.

DCAT 2 is adopted in FAIRification of Citizen Science platform \cite{10.1007/978-3-031-09917-5_34}, and open source platforms such as SEEK \cite{DBLP:conf/medinfo/LobeUBBBSIW21} to improve interoperability between digital assets on the Web and enable cross-domain markup. It is a core building block for developing REST API aiming at creating, storing, and serving FAIR metadata (see  FAIR Data Point (FDP) \cite{10.1162/dint_a_00160}).

DCAT is recommended by the ExPaNDS project as part of its \say{Final Recommendations for FAIR Photon and Neutron Data Management}\footnote{\url{https://doi.org/10.5281/zenodo.6821676} (accessed 17th February 2023)}.

\section{Conclusion and Future work}
\label{sec:conclusions}

DCAT 2 is a metadata schema that facilitates data catalogs' interoperability on the Web. 
%
DCAT gives people and machines a specific and domain-independent approach to create catalogs that express the core elements of a dataset description in a standardized way that is suitable for publication on the Web, and enables cross-domain interoperability by being used either on its own or alongside, as a complement to other data catalog standards. Thanks to this, DCAT facilitates effective search and retrieval and permits easy scaling up of the query process either through \say{frictionless} aggregation of dataset descriptions and catalog records from many different sources and domains, or by applying the same query across multiple catalogs and aggregating the results. These patterns can also be varied slightly so as to provide communities with tailored approaches to the dataset catalog that respect the specific nuances of a particular type of data.

DCAT 2 is designed as a community effort by DXWG, adheres to design principles specifically suited to establish it as a lingua franca for exchanging data coming from different catalogs. In particular,  the back compatibility with the previous version aims at preserving existing implementations; the reuse of terms from consolidated metadata vocabularies eases the interoperability promoting the adoption of cross-vocabulary modeling patterns; the minimization of ontology commitment opens to its reuse and specialization from the different domain communities; the Open-World Assumption unlocks DCAT complementation with other existing metadata vocabularies.

Version 2 builds on the initial work published in 2014 by providing, among other things, classes of descriptors that can be used for data services, and a wider set of relationships characterizing datasets and their temporal and spatial aspects. It also removes the constraints that were inherent in the prescribed use of some vocabulary terms for relationships (properties) that were present in its original version, so making their usage pattern more flexible. 

DCAT editors and DXWG support DCAT 2 adopters by assisting the specific doubts and issues via the DXWG public mailing list\footnote{\url{https://lists.w3.org/Archives/Public/public-dxwg-wg/} (accessed 06 March 2023)} and related GitHub space\footnote{\url{https://github.com/w3c/dxwg} (accessed 06 March 2023)}.
Further DCAT releases are planned, DXWG  is discussing including a more explicit notion of data series and versioning in DCAT. Going forward, the WG expects the incorporation of classes to describe data services into the model will make DCAT an increasingly useful tool in data science and provide a well-trodden path for those implementing the FAIR principles to follow.

\section*{Acknowledgement}
The authors gratefully acknowledge the contributions made to DCAT version 2 by all members of the working group, especially Annette Greiner, Antoine Isaac, Armin Haller, Dan Brickley, Ine de Visser, Jaroslav Pullmann, Lars G. Svensson, Linda van den Brink, Makx Dekkers, Nicholas Car, Rob Atkinson, Tom Baker.

The authors also gratefully acknowledge the chairs of the Data eXchange Working Group: Karen Coyle, Caroline Burle and Peter Winstanley — and W3C staff contacts Philippe Le Hégaret, Phil Archer and Dave Raggett.

Riccardo Albertoni was partially supported by TAILOR, a project funded by EU Horizon 2020 research and innovation programme under GA No 952215.

David Browning's work on this was funded by refinitiv.com (previously Thomson Reuters).




\bibliographystyle{elsarticle-num-names}
\bibliography{dcat2-vocabulary.bib}







\end{document}